\documentclass[manuscript]{acmart}

\usepackage[T1]{fontenc}

\usepackage{graphicx} 
\usepackage{color,soul}
\usepackage{paralist}
\usepackage{wrapfig}
\usepackage{tabularx}
\usepackage{multirow}
\usepackage{booktabs}
\usepackage{subcaption}
\usepackage{xcolor}
\usepackage{colortbl}
\usepackage{fontawesome5}
\usepackage{balance}
\usepackage{framed}
\usepackage{longtable}
\usepackage{url}
\usepackage{setspace}

\everypar{\looseness=-1}
\newcommand\gencite[3]{\cite[#3]{#1}}
\newcommand\genbibitem[3]{\bibitem{#1} #3}

\newcommand\fRole{1}
\newcommand\fResponsible{2}
\newcommand\fClear{3}
\newcommand\fAccurate{4}
\newcommand\fPersonal{5}
\newcommand\fMisuse{6}
\newcommand\fMitigations{7}
\newcommand\fReporting{8}
\newcommand\fEight{9}

\newcommand\highl[2]{\hl{#1} \textcircled{#2}}

\newcommand\tplot[1]{\raisebox{-.3\height}{\includegraphics[scale=0.135]{#1}}}

\setcopyright{none}

\newcommand\figureA[0]{
\begin{wraptable}{r}{0.62\textwidth}
\caption{Compliance with selected policy requirements}
\label{tab:compliance}
\footnotesize
\begin{tabular}{ll}
\toprule
Policy requirement & \% Compliance \\\midrule
\textcircled\fRole\ System description \& model role & \tplot{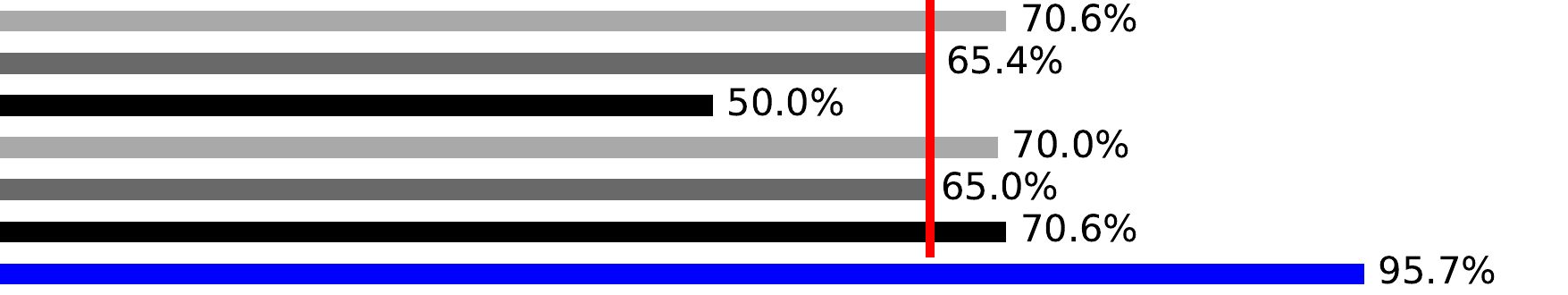} \\[.6em]
\textcircled\fResponsible\ Organization responsible & \tplot{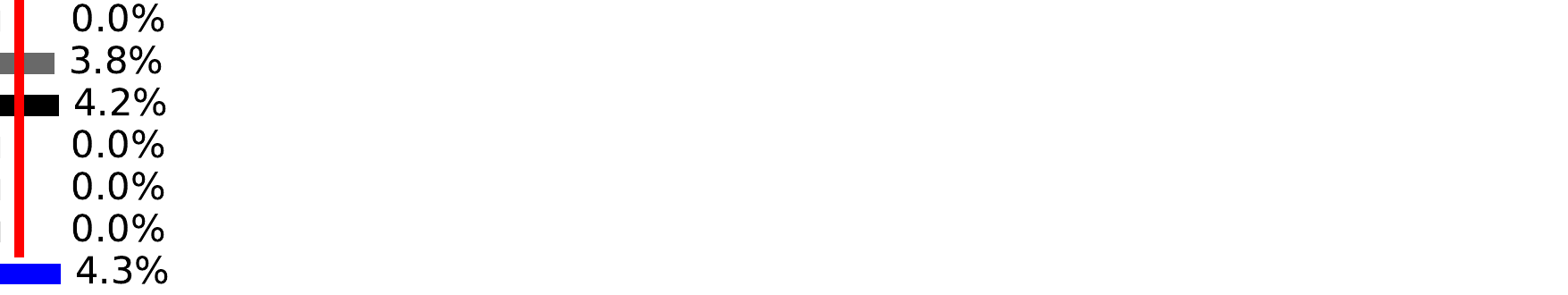} \\[.6em]
\textcircled\fClear\ Clear and accessible for end users & \tplot{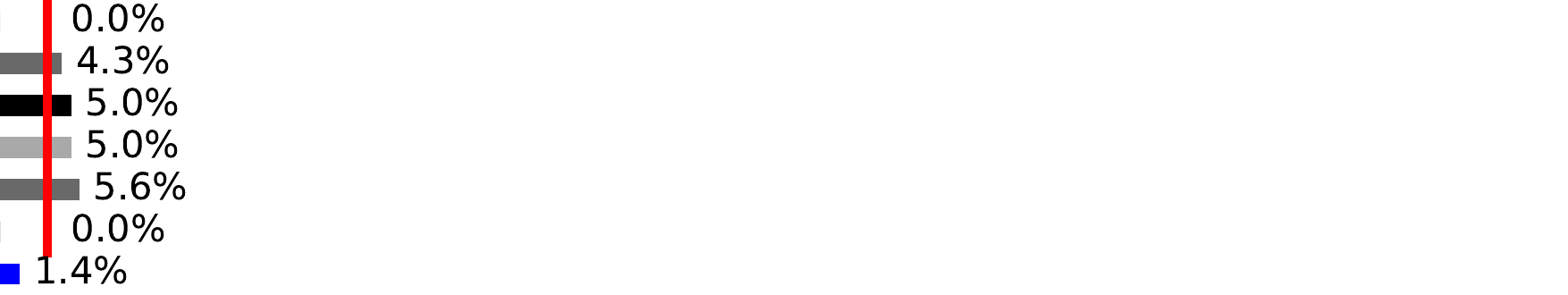} \\[.6em]
\textcircled\fAccurate\ Model accuracy & \tplot{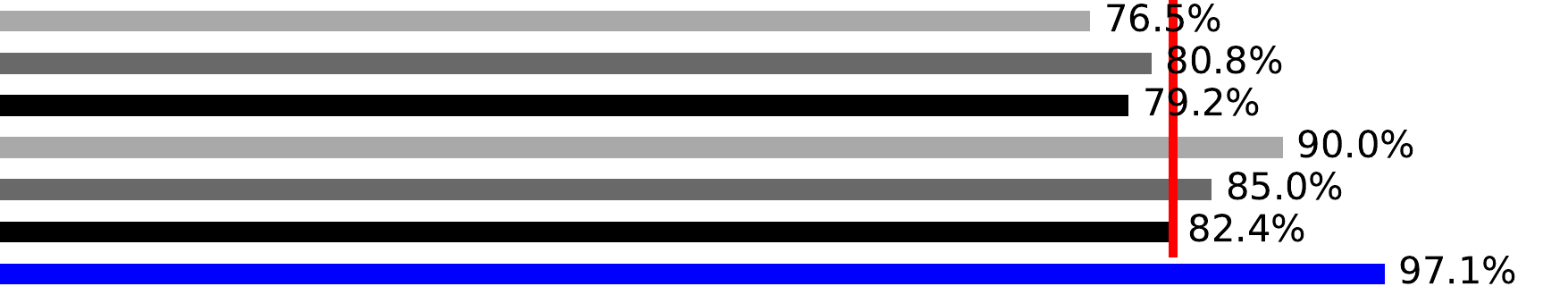} \\[.6em]
\textcircled\fPersonal\ Personal data used in prediction  & \tplot{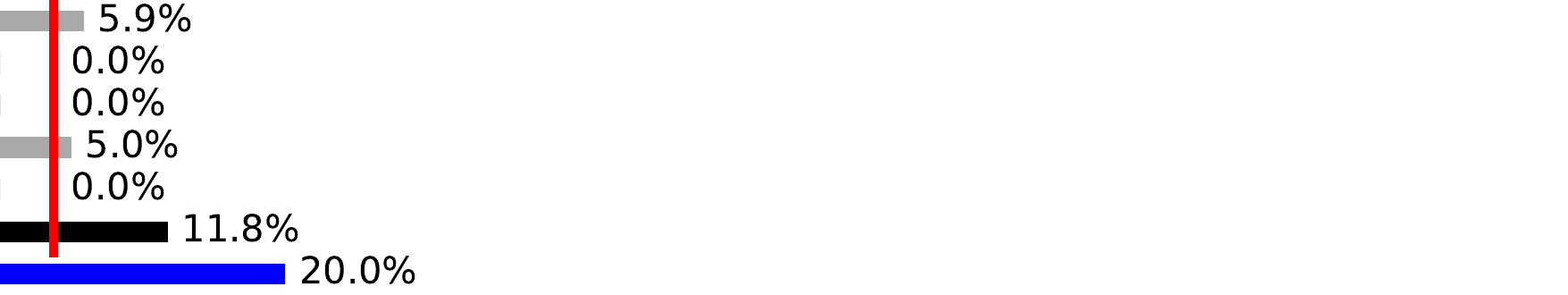} \\[.6em]
\textcircled\fMisuse\ Limitations and misuse & \tplot{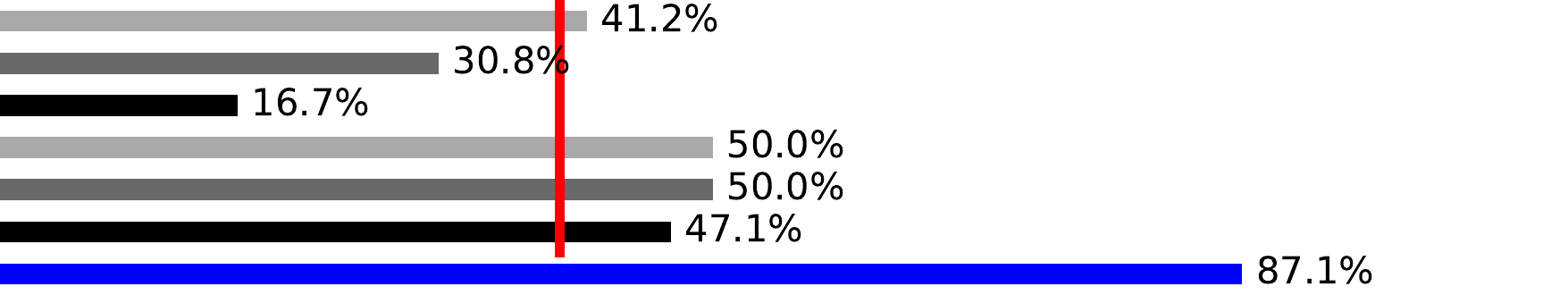} \\[.6em]
\textcircled\fMitigations\ Mitigations of limitations & \tplot{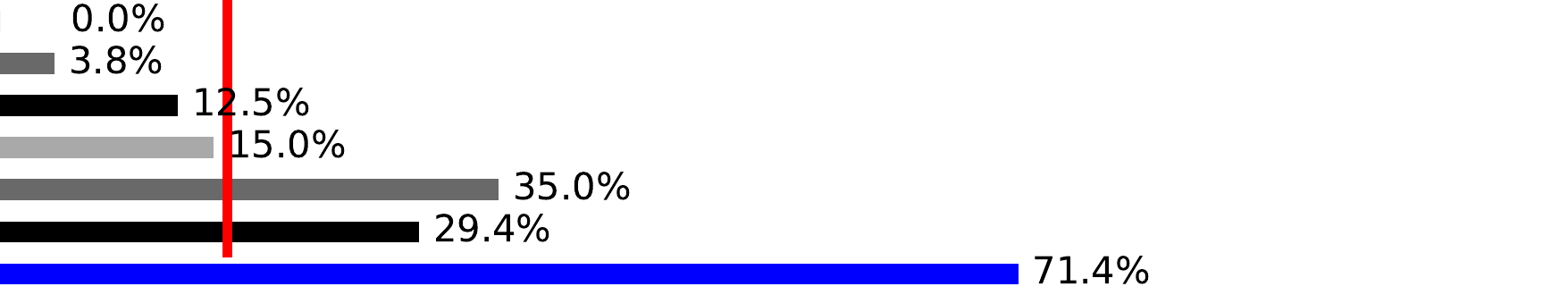} \\[.6em]
\textcircled\fReporting\ Reporting of misuse and harm & \tplot{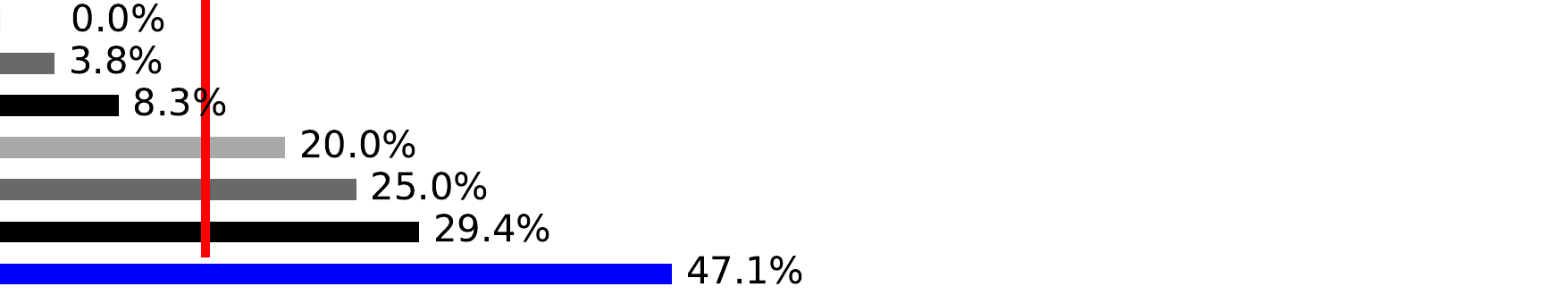} \\[.6em]
\textcircled\fEight\ Eight grade reading level\ & \tplot{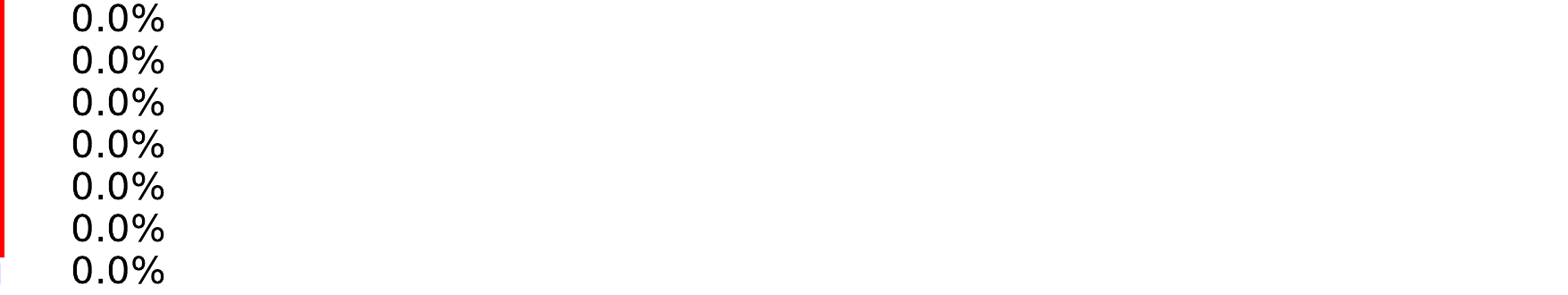} \\
\bottomrule
    \end{tabular}\\
{\scriptsize \textit{\begin{spacing}{0.9}
  Compliance in all six experimental conditions in experiment 1, from top to bottom: No purpose/short, dignity/short, human-AI col./short, no purpose/comprehensive, dignity/comprehensive, human-AI col./comprehensive. The vertical line indicates the average across all conditions. The blue line is for compliance in experiment 2. \\
\textcircled{\fRole}--\textcircled{\fClear} are included in the short policy; \textcircled{\fAccurate}--\textcircled{\fReporting} are included only in the comprehensive policy, \textcircled{\fRole}, \textcircled{\fClear}--\textcircled{\fEight} are enforced through grading rubrics in experiment 2.
\end{spacing}  }}
\end{wraptable}
}

\newcommand\tableCounts[0]{
\begin{wraptable}{r}{0.6\textwidth}
\centering
\vspace{-1em}
\caption{Experimental conditions and participant counts (n)}
\label{tab:conditions}
\vspace{-1em}
\footnotesize
\label{tab:participantnr}
\begin{tabular}{l l l l}
\toprule
\textbf{Study} & \textbf{Policy purpose} & \textbf{Policy length} & \textbf{$n$} \\
\midrule
\multirow{3}{*}{\parbox{3cm}{\textbf{Experiment 1} \\ \scriptsize\textit{Not enforced;} \\ \scriptsize\textit{participants self-selected} \\ \scriptsize\textit{stakeholder}}} 
& No purpose specified & Short & 17 \\
&                      & Comprehensive & 20 \\
& Human--AI collaboration & Short & 24 \\
&                      & Comprehensive & 17 \\
& Preserving dignity & Short & 26 \\
&                      & Comprehensive & 20 \\
\multirow{1}{*}{\parbox{3cm}{\textbf{Experiment 2} \\ \scriptsize\textit{Enforced; stakeholder provided}}}
& All conditions combined & & 70 \\
\vspace{0.1em}\\
\bottomrule
\end{tabular}
\vspace{-2em}
\end{wraptable}
}

\newcommand\figureFailure[0]{
\begin{wraptable}{r}{0.65\textwidth}
\centering
\caption{Failure mode comparison by explanation type}
\label{tab:failure-modes-detailed}
\footnotesize
\begin{tabular}{lll}
\toprule
Failure mode & Individual explanations & Global explanations \\
\midrule
1: \faRobot\ Inscrutable
& \tplot{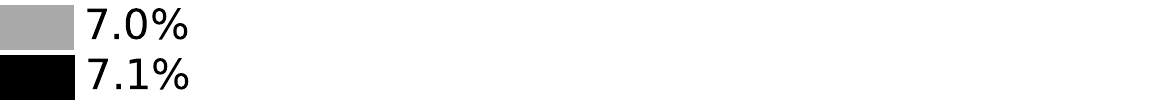}
& \tplot{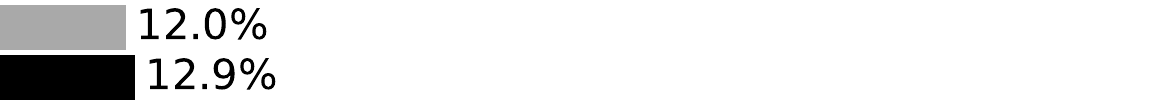} \\[.6em]

2: \faUserCog\ Requires ML expertise
& \tplot{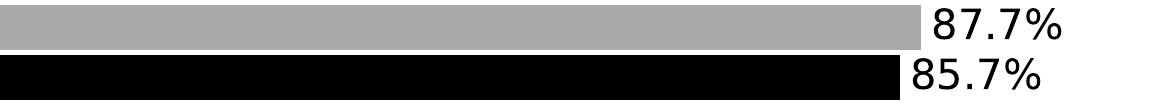}
& \tplot{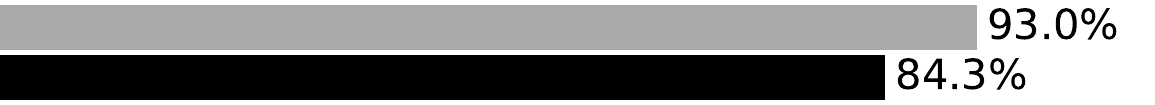} \\[.6em]

3: \faUserMd\ Requires medical expertise
& \tplot{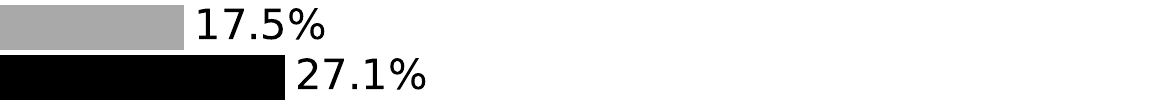}
& \scriptsize\textit{Not applicable}\\[.6em]

4: \faProjectDiagram\ Model-centric
& \tplot{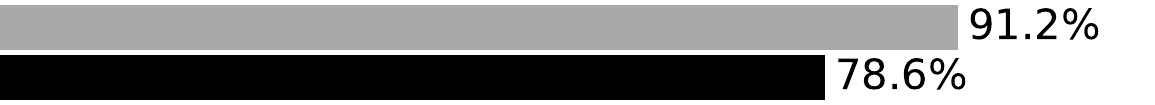}
& \tplot{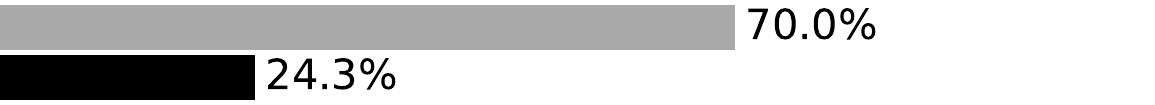} \\
\bottomrule
\end{tabular}\\

{\scriptsize
For each plot, the top bars correspond to experiment~1 and the bottom bars to experiment~2.
}
\vspace{-1em}

\end{wraptable}
}

\newcommand\figureB[0] {
\begin{figure}[t]
 \footnotesize
    \raggedright
    \textbf{Purpose of Policy:} To preserve the dignity of individuals | To enable effective human-AI collaboration | None \\
    \textbf{Policy Requirements:}
Designers, developers, and deployers of automated systems should provide generally accessible plain language documentation including \highl{clear descriptions of the overall system functioning and the role automation plays}{\fRole}, notice that such systems are in use, \highl{the individual or organization responsible for the system}{\fResponsible}, and explanations of outcomes that are \highl{clear, timely, and accessible}{\fClear}.\\
\textbf{Specifically: [comprehensive policy version only]}\\
INTENDED USE \\
\begin{itemize}
    \item Describe the automated system’s \highl{intended use and the role of the automation (model)}{\fRole}.  
    \item \highl{Provide evidence that the automation (model) functions accurately, consistently, and effectively in the intended use case}{\fAccurate}. 
\end{itemize}
HOW IT WORKS \\
\begin{itemize}
    \item Describe how the automation (model) works generally. Provide evidence that the documentation is effective for the policy purpose. 
    \item Provide a mechanism to describe how the automation (model) worked with regard to an instance of use to all intended users and subjects affected by the automated system \highl{in a form that is accessible to them}{\fClear}. Descriptions must include (1) that automation was used, (2) a short explanation of how the automation works, (3) what additional actors are involved in decisions, (4) \highl{what significant personal data was used for the decision}{\fPersonal}, (5) what decisions were reached in a specific case. Provide evidence that the documentation is effective for the policy purpose. 
\end{itemize}
CONCERNS\\
\begin{itemize}
    \item \highl{Describe limitations and misuse potential}{\fMisuse} of the automated system beyond its intended purpose and \highl{any provided mitigations}{\fMitigations}. 
    \item Describe the data used by the automated system. Justify the use of personal identifiable information. \\
    \item \highl{Describe how to report misuse}{\fReporting} or harm from the automated system.\\
\end{itemize}
LANGUAGE REQUIREMENTS\\
\begin{itemize}
    \item \highl{Provide all documentation in language appropriate for the intended audience. All documentation for untrained users must use nontechnical language at an eighth grade reading level}{\fEight}.
\end{itemize}
 \vskip -1em
        \caption{Our policy, highlighting the policy requirements selected for analysis (\textcircled\fRole--\textcircled\fEight)}
    \label{fig:policy}
 \vskip -2em
\end{figure}
}
\newcommand\figureOverview[0] {
\begin{figure*}[!h]
\includegraphics[width=1.1\linewidth]{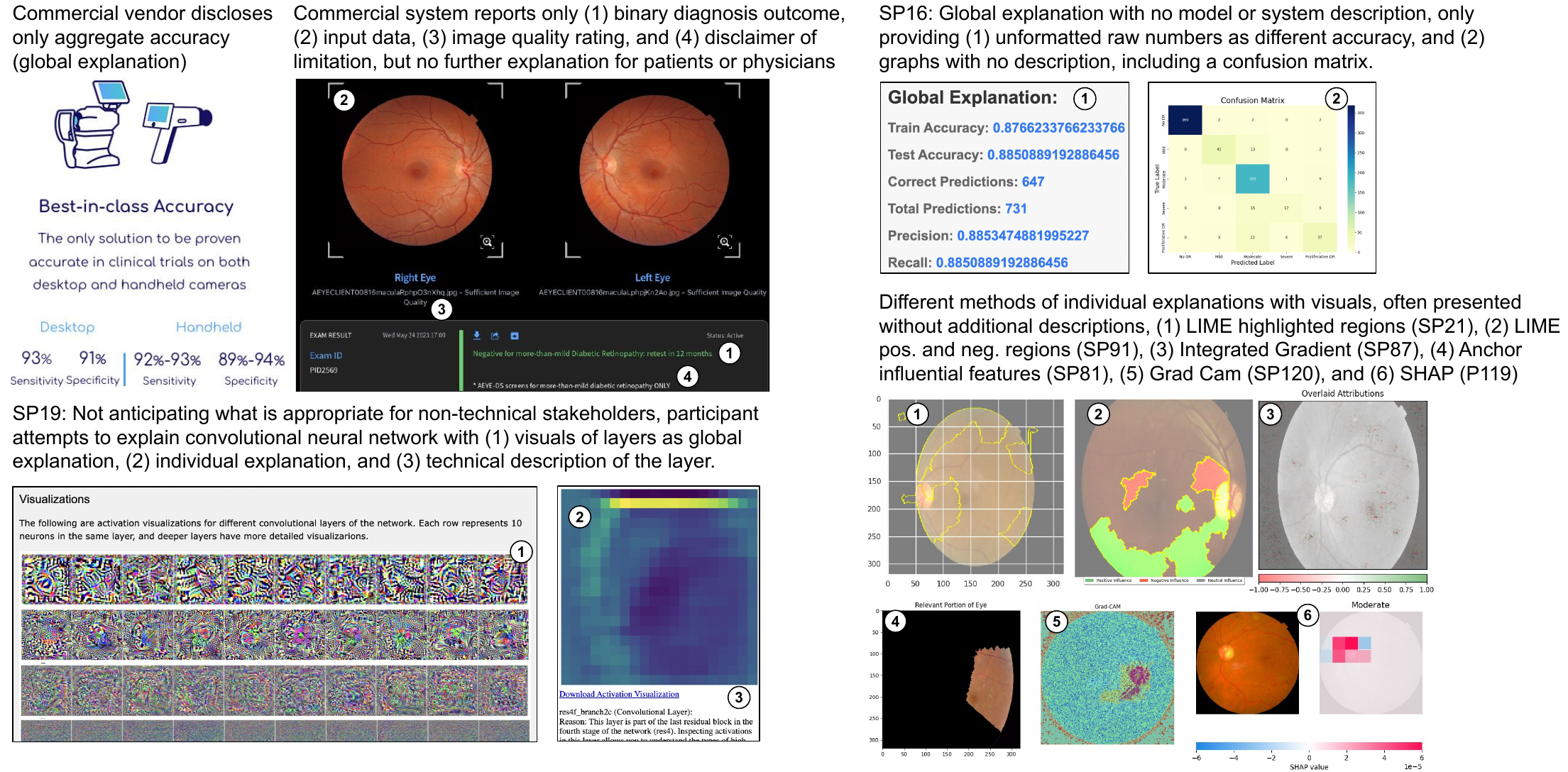}
\caption{Examples of explanations in commercial products and student solutions for diabetic retinopathy diagnosis}
\label{fig:overview}
\end{figure*}
}

\begin{document}
\looseness=-1
\title[Policy alone is probably not the solution: A large-scale experiment on how developers struggle to design ...]{Policy alone is probably not the solution: A large-scale experiment on how developers struggle to design meaningful end-user explanations}
\author{Nadia Nahar}
\authornotemark[1]
\email{nadian@andrew.cmu.edu}
\affiliation{%
   \institution{Carnegie Mellon University}
   \city{Pittsburgh}
   \state{PA}
   \country{USA}}
\author{Zahra Abba Omar}
\authornote{Both authors contributed equally to the paper}
\email{zahra.abbaomar@yale.edu}
\affiliation{%
   \institution{Yale University}
   \city{New Haven}
   \state{CT}
   \country{USA}}
\author{Jacob Tjaden}
\email{jay.tjaden@gmail.com}
\affiliation{%
   \institution{Colby College}
   \city{Waterville}
   \state{ME}
   \country{USA}}
\author{Inès M. Gilles}
\email{ines.gilles@yale.edu}
\affiliation{%
   \institution{Yale University}
   \city{New Haven}
   \state{CT}
   \country{USA}}
\author{Fikir Mekonnen}
\email{fikir.mekonnen@yale.edu}
\affiliation{%
   \institution{Yale University}
   \city{New Haven}
   \state{CT}
   \country{USA}}
\author{Erica Okeh}
\email{erica.okeh@bison.howard.edu}
\affiliation{%
   \institution{Howard University}
   \state{Washington, D.C.}
   \country{USA}}
\author{Jane Hsieh}
\email{jhsieh2@andrew.cmu.edu}
\affiliation{%
   \institution{Carnegie Mellon University}
   \city{Pittsburgh}
   \state{PA}
   \country{USA}}
   \author{Christian Kästner}
\email{kaestner@cs.cmu.edu}
\affiliation{%
   \institution{Carnegie Mellon University}
   \city{Pittsburgh}
   \state{PA}
   \country{USA}}
\author{Alka Menon}
\email{alka.menon@yale.edu}
\affiliation{%
   \institution{Department of Sociology, Yale University}
   \city{New Haven}
   \state{CT}
   \country{USA}}
\renewcommand{\shortauthors}{N. Nahar \& Z.A. Omar et al.}

\begin{abstract}
        Developers play a central role in determining how machine learning systems are explained in practice, yet they are rarely trained to design explanations for non-technical audiences. Despite this, transparency and explainability requirements are increasingly codified in regulation and organizational policy. It remains unclear how such policies influence developer behavior or the quality of the explanations they produce. We report results from two controlled experiments with 194 participants, typical developers without specialized training in human-centered explainable AI, who designed explanations for an ML-powered diabetic retinopathy screening tool. In the first experiment, differences in policy purpose and level of detail had little effect: policy guidance was often ignored and explanation quality remained low. In the second experiment, stronger enforcement increased formal compliance, but explanations largely remained poorly suited to medical professionals and patients. We further observed that across both experiments, developers repeatedly produced explanations that were technically flawed or difficult to interpret, framed for developers rather than end users, reliant on medical jargon, or insufficiently grounded in the clinical decision context and workflow, with developer-centric framing being the most prevalent.  These findings suggest that policy and policy enforcement alone are insufficient to produce meaningful end-user explanations and that responsible AI frameworks may overestimate developers' ability to translate high-level requirements into human-centered designs without additional training, tools, or implementation support. 
\end{abstract}
                \maketitle

        \section{Introduction}\label{h.l0c99u37w4ud}
\figureOverview

By now, it is broadly known that it is difficult to understand the internals of modern ML models. Many developers have used explanation techniques, such as LIME \gencite{kfNH}{(Ribeiro et al. 2016)}{} and SHAP \gencite{Dtgo,6RRB,sWKT}{(Lundberg and Lee 2017; Bhatt et al. 2020; Kaur et al. 2020)}{}, for debugging models and their predictions. More broadly, explainability and transparency are often seen as core responsible engineering practices that can help end users understand, collaborate with, oversee, audit, or contest systems with AI components \gencite{yO9q,aXCk,bJJc,URdB,pjxS,NCQQ,GbYG,WpoA}{(Selbst and Barocas 2018; Vera Liao and Varshney 2021; Rong et al. 2022; Cai et al. 2019; Nahar et al. 2024; Jacovi et al. 2021; Zhang et al. 2022; Luo and Specia 2024)}{}. For example, Holzinger et al. \gencite{AhJm7}{(Holzinger et al. 2019)}{} argue that explainability is the answer to ensuring greater use of ML-powered systems in healthcare: If healthcare providers can understand how a decision was reached, then reflecting on the output of an ML model is like any other screening or diagnostic tool. Explanations are also increasingly positioned as mechanisms for safety oversight and accountability, supporting activities such as auditing, error investigation, and post-deployment monitoring \gencite{yO9q,NCQQ}{(Selbst and Barocas 2018; Jacovi et al. 2021)}{}. By making potential failure modes visible, explanations can enable human oversight and intervention before harm occurs.

Designing ML-powered systems to be explainable and transparent to end users is challenging. A whole community of researchers and a small number of practitioners (often with a user-experience design specialization) is exploring how to design and evaluate effective end-user explanations under the label of \emph{human-centered explainable AI} \gencite{aXCk,52dm9,URdB,bJJc,FoP5}{(Vera Liao and Varshney 2021; Panigutti, Beretta, et al. 2023; Cai et al. 2019; Rong et al. 2022; Google PAIR 2019)}{} -- with many studies on various systems (and conflicting evidence). These efforts have revealed numerous strategies and common pitfalls. As yet, no replicable or scale-able solution has emerged: creating end-user explanations is still more like research than following a standard recipe, requiring careful consideration of target users, personas, and context, and often needing design support rather than step-by-step instructions. However, in contrast to model-focused technical explainability tools \gencite{QIqH}{(Molnar 2020)}{}, these topics are not broadly taught nor as easily usable as easily installed tools or libraries, and it is likely that few developers have encountered them. In practice, most projects do not have access to experts with research experience or dedicated training in human-centered explainable AI.

In a series of two controlled experiments \emph{with 194 participants} (encompassing about 1,552 hours of work in total, in a graduate level course covering software engineering, machine learning, and MLOps), we explored how developers, without dedicated training in human-centered explainable AI, design end-user explanations for an ML-powered medical device, and to what degree policy guidance can shape their behaviors toward responsible practices and effective explanations. We explore policy guidance, because such guidance from laws \gencite{LqUJ}{(“Press Releases: Artificial Intelligence Act: MEPs Adopt Landmark Law” 2024)}{}, from in-house policies in corporations \gencite{xN9H,66ix}{(Microsoft 2022; Jobin et al. 2019)}{}, or from professional organizations \gencite{cYOq,FSB8}{(Technology Policy Committee 2024; Chatila and Havens 2019)}{} is often seen as a potential tool to shape behaviors in lower-risk applications.  Although historically, regulation is focused on high-risk applications (e.g., aviation, healthcare), it may provide a path to instill responsible engineering practices more broadly. Many jurisdictions have explored or are exploring AI regulation \gencite{oSHb,LqUJ}{(U.S. White House 2022; “Press Releases: Artificial Intelligence Act: MEPs Adopt Landmark Law” 2024)}{}, and many corporations publish their own responsible AI policies with different processes attached \gencite{xN9H,66ix}{(Microsoft 2022; Jobin et al. 2019)}{}.

We argue that understanding and shaping developer behavior broadly is a promising path toward more responsible AI products, as it is often developers, with deep knowledge in their own field of data science or software engineering, who make important and consequential decisions with little oversight. Identifying effective means to change their behavior toward responsible engineering practices can provide strong leverage to improve software products and avoid harms to their users.

With our experiments, we asked the following research questions:

\begin{compactitem}
	\item RQ0: What explanations do developers design for end users?

	\item RQ1: How do differences in policy detail and policy purpose influence policy compliance and quality of developer explanations when policy is provided as guidance?

	\item RQ2: How does policy enforcement influence policy compliance and quality of developer explanations?

\end{compactitem}

In a nutshell, our experiments found that (a) across all experimental conditions, most developers in our experiment failed to take the end-users' perspectives and instead provided explanations that would be inscrutable to intended end users like nurses and patients, and (b) that policy enforcement improved compliance with the policy (to a degree) but did little to improve explanation quality. Participants were mostly proficient in using libraries to produce technical explanations, but often failed to consider the context of how explanations would be used in a practical setting. While policy guidance and enforcement changed some behaviors in a somewhat mechanical way, it did not lead to learning and deeper engagement with the core purpose.

Cynically speaking, we could argue that our experiment merely confirms the common trope that computer science students or developers lack empathy for users and that we need to involve requirements engineers and UX designers for a reason -- mirroring  observations in the Alan Cooper's \emph{``The Inmates are Running the Asylum''} book \gencite{hptf}{(Cooper 2004)}{}. We could also argue that providing the policy without additional training, implementation guides, or institutional support was doomed to fail, given that established regulatory frameworks like FDA, DO-178C, and Common Criteria typically rely on layered organizational control structures, experts, and consultants \gencite{MD78,d5Ve}{(Ferreira et al. 2019; Papademetris et al. 2022)}{}. Nevertheless, we argue that our experiments provide a useful data point in two ways:

\begin{compactitem}
	\item Our experiment provides clear evidence that developers should not be expected to design end-user ML explanations without additional training. 

	\item Our experiments dampen the optimism that policy might be a broad, lightweight, and effective intervention for responsible AI: Ambitious policy documents like the White House Blueprint for an AI Bill of Rights \gencite{oSHb}{(U.S. White House 2022)}{}, espoused responsible AI principles by many companies big and small  \gencite{xN9H,66ix}{(Microsoft 2022; Jobin et al. 2019)}{}, and even the EU AI Act  \gencite{LqUJ}{(“Press Releases: Artificial Intelligence Act: MEPs Adopt Landmark Law” 2024)}{}, are light on implementation details. Our experiment shows that policy alone is not a magic shortcut to get developers to design better and more responsible explanations, while traditional intensive and expensive regulatory frameworks on medical devices and aviation will be difficult to scale to everyday ML-powered applications. Because these still have substantial potential for harm \gencite{JBqy,c581,WtvE,JHix,dFKX,OdxV0,siRJO,gbBY9,FmF2E}{(Raji et al. 2022; Weidinger et al. 2021; O’Neil 2016b; Barocas and Selbst 2016; Kästner 2025; Eubanks 2018; Fourcade and Healy 2013; Gandy 1993; Noble 2018)}{}, it is important to fill this gap.

\end{compactitem}

Our study also provides a starting point for other incremental interventions. Understanding the specific failures observed in our experiment (including developers' failure of imagination and their check the box compliance), future work can now work on more targeted interventions, while still keeping the overall process lightweight. Interventions are urgently needed, whether better training, better tooling, or better implementation guidance of policy guidelines.

In summary, we contribute (a) results from two large-scale controlled experiments on how developers (fail to) design meaningful end-user explanations and (b) a discussion of explanation problems and pathways for improvement that can be taken up by policymakers and educators.

\section{Background and related work}\label{h.2hticebpn86y}
Machine learning components (from traditional ML to LLMs to agents) are increasingly integrated into software products, where they produce outputs, suggest decisions, or even automate actions in the real world \gencite{yj4R,dFKX,J76x}{(Hulten 2019; Kästner 2025; Amershi, Begel, et al. 2019)}{}; this includes medical devices and diagnostic tools promising  lower costs and improved health outcomes \gencite{Lalin}{(Chin-Yee and Upshur 2019)}{}. Modern ML models are usually complex and inscrutable, even to their creators, where developers cannot simply inspect model internals to understand how exactly the model works. Software engineers who want to ensure the quality of the overall software product (including the safety of a medical device) hence need to understand how to integrate ML components and how to compensate for their inaccuracies, possibly through safeguards around the model \gencite{dFKX,7G0Kn,yj4R,hKo4m}{(Kästner 2025; Dong et al. 2024; Hulten 2019; Costa et al. 2025)}{} and human-computer-interactions design \gencite{FoP5,m95Zv,c7Fqz}{(Google PAIR 2019; Yang et al. 2020; Amershi, Weld, et al. 2019)}{}.

Without insight into inner workings of a model, developers risk building systems that are unreliable, biased, misleading, or manipulative \gencite{vmOf,yO9q,URdB,2310}{(Rudin 2019; Selbst and Barocas 2018; Cai et al. 2019; Springer et al. 2018)}{}. Users may have difficulty trusting, overseeing, and effectively working with an ML-powered system, failing to correct mistakes, such as an obviously wrong diagnosis from medical software. \emph{Explainability} is multi-faceted and can serve many purposes \gencite{QIqH,jG9Nd,yO9q}{(Molnar 2020; Lipton 2018; Selbst and Barocas 2018)}{}.   \emph{Explanations} as communication made by humans to other humans (e.g., a doctor explaining a diagnosis) provide a possible analog for ML explainability: Rather than explaining every step in an explicit algorithm, they provide a necessarily partial, approximated explanation, targeted to the needs of the recipient \gencite{bwS3Y,p8Hx}{(Esposito 2023; Menon et al. 2024)}{}. \emph{Global} explanations aim to explain the overall behavior of a model, e.g., with partial dependence plots and feature importance (Molnar, 2020); whereas \emph{individual} explanations provide information about how the model arrived at a specific output for a given input (e.g., a medical prognosis), e.g., identifying influential features with SHAP \gencite{Dtgo,6RRB,QIqH}{(Lundberg and Lee 2017; Bhatt et al. 2020; Molnar 2020)}{}. Currently, these techniques are mostly used \emph{by developers} to debug model behavior \gencite{6RRB,sWKT}{(Bhatt et al. 2020; Kaur et al. 2020)}{}.

We consider also broader considerations of \emph{transparency} (a term common in AI policy language \gencite{66ix,uU4a,AX2E}{(Jobin et al. 2019; Panigutti, Hamon, et al. 2023; Nannini et al. 2023)}{}) as part of explainability, such as explaining that a model is used in the first place, what the model is used for, what personal data is used and why, and whether there is a path to appeal an automated decision \gencite{uU4a}{(Panigutti, Hamon, et al. 2023)}{}. When asked about explanations (e.g., in co-design studies \gencite{CkiE5}{(Luria 2023)}{} and our own research), end users tend to express that they do not desire detailed technical explanations, assuming they would not understand them; instead, they prioritize information about the model's existence, the data used, and audits performed by third-parties. 

Explanations are usually intended to serve a purpose\emph{,} whether functional, social, economic, or normative \gencite{pjxS,yO9q,4Wic,bJJc,Tu1qh}{(Nahar et al. 2024; Selbst and Barocas 2018; Colaner 2022; Rong et al. 2022; Alpsancar et al. 2024)}{}, but that purpose is rarely articulated clearly in discussions, requirements, or even regulation. Beyond debugging, purposes include (1) \emph{auditing}, especially for fairness issues \gencite{yO9q,GbYG}{(Selbst and Barocas 2018; Zhang et al. 2022)}{}, (2) \emph{human-AI collaboration} for effective use and calibrating trust \gencite{WpoA,URdB,FoP5}{(Luo and Specia 2024; Cai et al. 2019; Google PAIR 2019)}{}, (3) \emph{oversight} and \emph{contestation} of wrong data and decisions \gencite{yO9q,ZSyX}{(Selbst and Barocas 2018; Wachter et al. 2017)}{}, and (4) assuring the \emph{dignity} and agency of individuals  \gencite{4Wic,yO9q}{(Colaner 2022; Selbst and Barocas 2018)}{}. For many of these purposes, explanations must be aimed at end users or external parties, not just developers.

Critiques of a lack of end-user focus go back to the early days of explainability research \gencite{lmS4}{(Miller et al. 2017)}{} and lead to the emergence of the \emph{human-centered explainable AI} community \gencite{aXCk,52dm9,URdB,bJJc,FoP5}{(Vera Liao and Varshney 2021; Panigutti, Beretta, et al. 2023; Cai et al. 2019; Rong et al. 2022; Google PAIR 2019)}{}, which  explores designing effective explanations for \emph{end users}, e.g., to improve human-AI collaboration. However, end-user explanations are generally less studied and less deployed than technical explanations for developers, and evidence for effectiveness is mixed \gencite{bJJc}{(Rong et al. 2022)}{}. Many studies highlight risks for manipulation of user behavior through explanations, e.g., \gencite{KZKub,8FqpS,yRYe0}{(Eiband et al. 2019; Stumpf et al. 2016; Ehsan et al. 2021)}{}, and recognize that explainability needs for end users are context-dependent beyond one-size-fits-all solutions \gencite{mZTPq}{(Laato et al. 2022)}{}.

Regulation provides a form of societal infrastructure for coordinating social welfare and distributing risks, and establishing paths toward standards of practice \gencite{wUt8O}{(Marsden 2014)}{}. Sociological scholarship often makes a distinction between two types of relevant law in medical contexts \gencite{gIwn}{(Heimer 2025)}{}: ``Hard law'' is typically passed by legislative bodies and enforceable by action of regulatory agencies or in court, backed by the authority of the government; the EU AI Act is one relevant example \gencite{LqUJ}{(“Press Releases: Artificial Intelligence Act: MEPs Adopt Landmark Law” 2024)}{}. ``Soft law,'' by contrast, includes rules and guidelines that are written by a range of non-legislative bodies, as well as the guidance to implement and interpret hard law (still emerging in the case of the EU AI Act). Policy can have effects at two levels: (1) at a legalistic level, motivating changes to behavior to avoid a pre-specified sanction, like fines, and (2) at a normative level, indicating what is symbolically valued or desired and setting expectations for what constitutes good professional behavior. Even when it is not enforced or enforceable, policy can signal values and provide a basis for professionals with less power to challenge or shift the status quo \gencite{gIwn}{(Heimer 2025)}{}. Against a recent turn toward de-regulation for AI-powered in the U.S., it is all the more important to assess what is possible for ``softer'' policy guidance to achieve, particularly in an early phase where developers move fast with emerging technologies before norms emerge about responsible engineering practices. To date, policymakers have sketched out broad policy principles aimed at shaping developer behavior on transparency for ML systems, such as White House \emph{Blueprint for an AI Bill of Rights} \gencite{oSHb}{(U.S. White House 2022)}{} and former Executive Order 14110 \gencite{BaSe9}{(The White House 2023)}{}. Additionally, many companies have/are investing in in-house responsible AI internal guidelines and practices \gencite{xN9H,66ix}{(Microsoft 2022; Jobin et al. 2019)}{}. 

However, law in any form has important limitations. Evidence from social science suggests that organizations, institutions, and professional norms, in addition to law, play roles in changing the actions of professionals like developers \gencite{ueqM,GZkE,N0Av}{(Gray and Silbey 2014; Silbey 2013; Cech and Sherick 2015)}{}. At its most effective, policy provides clear guardrails that enable innovation \gencite{oT4B}{(Suran M 2024)}{} by balancing between competing demands: It must ensure an even playing field for technological development and commercial exchange while not creating so onerous a burden that innovation is stifled \gencite{0S7F}{(Nelson 2024)}{}. In practice, in high-risk domains (e.g., healthcare, aviation), regulation and policy are heavyweight and sometimes cumbersome, entailing substantial infrastructure, guidance, and consultants \gencite{MD78,d5Ve}{(Ferreira et al. 2019; Papademetris et al. 2022)}{}. By contrast, many AI guidelines aim to cover applications across many domains, taking a more lightweight approach. It remains a question what developers (without dedicated training or access to experts) make of this kind of policy guidance, and what other support might be necessary to ensure compliance, given context-specific and application-specific explanation needs \gencite{mZTPq}{(Laato et al. 2022)}{} and competing demands on developers' attention, like time pressure, conflicts of interest, and regulatory capture \gencite{9fBjx,dQM5t,bevHA,thHU8,Hgm76}{(Bietti 2020; Metcalf et al. 2019; Greene et al. 2019; Ochigame 2019; Ferretti 2022)}{}. Research on how developers receive, interpret, and enact guidelines--how, in short, they navigate between the legalistic and normative levels of law--is necessary to help better align policy and developers. 

\section{Study design}\label{h.r6f0el8l6rfk}
To explore how developers design end-user explanations generally (RQ 0) and how policy guidance (RQ 1) and policy enforcement (RQ 2) influences compliance and explanation quality, we conducted two controlled experiments in the context of a graduate course. Across both experiments, we tested 7 experimental conditions with different policy language and different degrees of enforcement with 194 participants.

\subsection{The scenario: Diabetic retinopathy screening}\label{h.mqiw84rqhn5}
Participants were asked to provide explanations for a hypothetical low-cost ML-powered medical device to screen for diabetic retinopathy. The device detects diabetic retinopathy on a scale of 0 to 4 (none to severe) using images of the eye and the patient's age and gender, comparable to existing commercial screening tools. The device would be used by trained users (e.g., nurses or volunteers) to perform screenings at mobile clinics or in patients' homes, with the potential, as stated in the scenario, to \emph{``drastically reduce screening costs and make screenings much more available, especially in under-resourced regions of the world.''} Related  (more costly) devices are commercially available \gencite{PFlIB,u9583,G1Bl9}{(Ng 2020, 2022; Hampton 2018)}{}; in Fig.~\ref{fig:overview}, we show the limited explanations for/by one of them. Existing rates of compliance with annual screening recommendations for diabetic retinopathy among diabetics in the U.S. range from 25 to 60\% \gencite{gKtFa}{(SK Ha, JB Gilbert, E Le, C Ross, and A Lorch 2025)}{}. 

We chose this scenario for its real-world application, current relevance, and readily available data and models. In preparation, we conducted interviews with regulators of medical devices, medical professionals, and diabetes patients, asking how they approached understanding screening device predictions, complying with clinical norms and regulations, and integrating tools into clinical practice. Over two years, we attended large diabetes conferences, where we interacted with representatives of companies (both startups and established firms) marketing ML-powered diabetic retinopathy screening devices and observed how screening tools were introduced to physicians. This preparation provided us with more background knowledge than most the average non-clinician researcher  to evaluate participant solutions from the perspective of clinical practitioners and patients.

\subsection{Tasks }\label{h.or5tuqwts5r}
We provided a dataset (from a public dataset used for a Kaggle competition \gencite{ZUG8U}{(“APTOS 2019 Blindness Detection” 2019)}{}) and a pre-trained ResNet50 model. We augmented the data with synthetically generated gender and age data to enable participants to perform segmented analysis of subpopulations and describe the use of potentially sensitive information. 

The task was to create explanations for the system that comply with a provided policy (see below), creating (HTML) pages that present: (a) \emph{Global explanations:} What external stakeholders might want to know about the product, the model, or the data. This might be information found on the product web page, training materials, or a handbook. (b) \emph{Individual explanations:} Information about a specific diagnosis. This might be shown on the device, recorded in the patient's medical records, or provided as a printed handout. 

In the first experiment, we asked participants to identify targeted stakeholders themselves;  in the second, we specified that the handbook was intended for nurses/volunteers and the handout for patients. In addition, we asked participants (a) to describe their solution, (b) to self-assess their compliance with their assigned policy and provide evidence of their compliance, and (c) to write a reflection about the challenges they faced. 

Participants were given basic training in explainability techniques and transparency as part of their coursework prior to completing the task (160 minutes of lectures, two readings \gencite{vmOf}{(Rudin 2019)}{} \gencite{FoP5}{(Google PAIR 2019, chap. 3)}{ch.~3}, and an 80 minute lab session); instructions briefly covered the pitfalls of explanations and the diverse needs of different stakeholders (using the ``Hello AI'' case study \gencite{URdB}{(Cai et al. 2019)}{}), but mostly focused on technical post-hoc explainability techniques like \emph{LIME} and \emph{Anchors} \gencite{QIqH}{(Molnar 2020)}{}. Participants were not given instruction about diabetic retinopathy or clinical communication. We designed the task to be about 8 hours of work per participant, not including prior training.

\figureB

\subsection{Experimental conditions (independent variables): Policy length, purpose, and enforcement}\label{h.g0x4ij6g1efq}
In the first experiment, we provided policy as guidance and required self-assessment but varied the \emph{comprehensiveness} of the policy and its provided purpose (6 conditions): We either provided a one-sentence policy extracted from the \emph{Blueprint for an AI Bill of Rights} \gencite{oSHb}{(U.S. White House 2022)}{} or a more comprehensive version that \emph{additionally} included a \emph{prescriptive list} of requirements, inspired by recent research on policy design \gencite{pjxS}{(Nahar et al. 2024)}{};   for each, we provided one of three stated \emph{purposes} of the policy as either (a) \emph{``to enable effective human-AI collaboration,''} (b) \emph{``to preserve the dignity of individuals,''} or (c) no purpose was stated. In Fig. \ref{fig:policy}, we show the text of all policy versions. After learning that policy differences had little influence in our first experiment, we conducted a second experiment, assigning the same policy (the comprehensive policy, without the initial sentence, with the purpose of effective human-AI collaboration for nurses and preserving dignity for patients) to all participants, but enforcing the policy through a grading rubric rather than asking for self-assessment only.

\subsection{Recruitment and participants}\label{h.84vzwf6uvg7k}
Participants were recruited from a large graduate course on software engineering, machine learning, and MLOps in two consecutive semesters [details omitted for anonymity]. In the course, most students already had substantial prior experience as software engineers or data scientists: 63\% had prior internship, research, or work experience as a data scientist, and 51\% had internship, research, or work experience as a software engineer, including 29\% of students who had previously worked in industry as a data scientist or software engineer (or both). Only 6\% and 5\% of students indicated having no prior data science or software engineering experience respectively. The students' background is reflective of many early-career practitioners in industry teams, who usually have experience in their field and basic awareness of explainability tools, but limited exposure to human-centered explainable AI. While they likely have personal experience with medical devices as patients, our participants were unlikely to have the domain expertise or the access to domain experts that would come with working in an industry team on a commercial product.

\tableCounts

The IRB approved study was designed as a secondary analysis of a homework assignment. All students in the course had to complete the homework assignment and were graded based on a standard rubric. In the first semester, the rubric did not require policy compliance and was orthogonal to the six experimental conditions; in the second semester, all students were given the same policy and were uniformly graded on compliance. Students could opt to allow us researchers to perform an analysis of the anonymized assignment after the submission of final grades at the end of the semester (194 did, 16 did not). Participants did not receive any monetary or credit incentives. In the first iteration, participants were randomly assigned to six groups, and in the second iteration all participants were assigned to the same \emph{policy-enforced} condition (see Table~\ref{tab:participantnr}). 

While we know the demographics of students in the course generally, we intentionally did not collect individual background information of participants due to research ethics considerations and to avoid raising barriers to participation. Random assignment of large experimental groups makes substantial experience/demographic differences among the groups unlikely. Demographics and experience were similar across both semesters.

\subsection{Data analysis (incl. dependent variables)}\label{h.etl2ayxrs229}
We analyzed all solutions using \emph{qualitative content analysis} \gencite{mqt5J}{(Schreier 2012)}{}, where researchers create coding rubrics for one or more dimensions and systematically assign one code per dimension to each chunk of analysis (here each participant's solution is considered as one chunk). Qualitative content analysis uses qualitative research methods for interpreting meanings, themes, and patterns within content through inductive reasoning and contextual understanding for systematic coding that \emph{produces frequency counts that can be analyzed quantitatively}. 

We analyzed the solutions regarding three research questions: For \emph{RQ 0}, we identified elements of explanations in terms of what form the explanations have (e.g., text, visuals), what data is presented (e.g., confusion matrix), and what post-hoc explanation tooling was used (e.g., SHAP). For \emph{RQ 1}, we judged policy compliance of each solution for nine specific policy requirements highlighted in Fig.~\ref{fig:policy}. We purposefully selected a subset of policy requirements to scope the analysis, including requirements related to global (e.g.,  \textcircled{\fRole}, \textcircled{\fResponsible}, \textcircled{\fMisuse}) and individual (e.g., \textcircled{\fEight}, \textcircled{\fPersonal})  explanations, requirements that require deep design (e.g., \textcircled{\fAccurate}) and requirements that are met with fact statements (e.g., \textcircled{\fResponsible}, \textcircled{\fReporting}). We assessed compliance with the requirement to write explanations at an 8th-grade reading level automatically through the common/standard/validated measure of \emph{Flesch-Kincaid (FK) Grade Level} \gencite{Adih,lfHY,E6i7}{(Williamson and Martin 2010; Challener et al. 2025; Tahir et al. 2020)}{}. For \emph{RQ 2}, independent of compliance, we coded for four common failure modes and corresponding symptoms that resulted in poor quality explanations discussed in detail below. For the first iteration where we left the choice of stakeholder to the participants, we only analyzed those solutions that targeted nurses for global explanations (n) and patients for individual explanations (n) to enable more meaningful comparisons. 

As standard for this method \gencite{mqt5J}{(Schreier 2012)}{}, the codebook was developed based on domain knowledge and an analysis of a subset of the solutions, before applying it to all solutions. Especially for RQ2, we first analyzed common problems in the solutions in an open-ended way (mirroring thematic analysis \gencite{OmP5p,ucbFD}{(Lazar et al. 2017; Onwuegbuzie and Burke Johnson 2021)}{}), settling on the coding frame only after many discussions, once we reached saturation. We share the codebook in the appendix \gencite{1jmMl}{(Appendix: Policy Alone Is Probably Not the Solution: A Large-Scale Experiment on How Developers Struggle to Design Meaningful End-User Explanations 2025)}{}. To increase reliability, after our initial manual coding, we repeated the coding for RQ2 with an LLM, and investigated every disagreement between the model and the original labeler (6\% to 31\% of labels per dimension), and corrected 92 labels out of 868. For RQ3, LLMs were unreliable and we instead assessed inter-rater reliability on 40 solutions (20 individual, 20 global explanations) with two raters, yielding Cohen's k values between 0.83 and 1, indicating almost perfect agreement. For the resulting quantitative data, we report descriptive statistics and refer the interested reader to the appendix for (often negative) statistical results from chi-squared tests and logistic regressions.

\subsection{Limitations and threats to validity}\label{h.jofw7hkx30o0}
As with every study, ours also has limitations from tradeoff decisions in the research design, and the results should be interpreted accordingly. First, we are an interdisciplinary research team from four US-based universities with distributed expertise in machine learning, software engineering, and social science. We have interacted with and interviewed manufacturers and users of diabetic retinopathy screening tools (see above), giving us more domain knowledge than the participants. Despite carefully calibrated rubrics and assessed inter-coder reliability, our backgrounds may bias us towards assessing explanations more critically than the average population of users. Second, conducting a study with graduate students has recognized benefits and drawbacks \gencite{BsX0Z,F1Nts,wZRFD}{(Salman et al. 2015; Falessi et al. 2018; Feldt et al. 2018)}{}. The classroom setting allowed us to conduct the study at a scale (number of participants and task length/depths) that would be infeasible with professional developers. With a majority of our participants having prior internship, research, or work experience, we regard them as representative of early-career professionals about to (re-)enter technology careers upon graduation. As their education is more recent and they were introduced to explainable AI through course content, participants might be more primed for responsible AI engineering than most practitioners. Participants may be biased to use techniques explicitly introduced in the course, but this is orthogonal to our RQ1 and RQ2 findings. In contrast, the typical practitioner would likely have more domain knowledge about healthcare. Readers should exercise care when generalizing results beyond our population. 

\section{Results}\label{h.t0g88yw22cjp}
We report results by research question, starting with general observations across all policy conditions, before analyzing differences among experimental groups.

\subsection{Participants provide mostly technical explanations with off-the-shelf tools (RQ 0)}\label{h.i2hs21knm0}
While the policy provided high-level guidance or requirements, the assignment did not prescribe \emph{how} to provide explanations. See Fig.~\ref{fig:overview} for illustrative excerpts of some solutions. In experiment 1, the majority of participants provided technical information about model evaluation and training for global explanations (e.g., SP16 in Fig.~\ref{fig:overview}), with over half reporting cohen kappa scores, confusion matrices, and description of training data distributions. Many participants (21\%) provided disaggregated evaluation results for subpopulations by age, gender, or severity. A few participants provided technical details of the model architecture (e.g., SP19 in Fig.~\ref{fig:overview}). About half of the solutions provide a description of the purpose of the model in the system. For individual explanations, almost all solutions (98\%) included a visual explanation highlighting pixels or overlaying boxes on the input image (63\% used anchors, 19\% LIME, 8\% SHAP, 10\% others; as in Fig.~\ref{fig:overview}), however often without any description on how to interpret the image. Generally, participants used explainability techniques that are readily available from libraries. Explanations in the second experiment were similar.. 

\subsection{Policy language barely influenced compliance, but enforcement did (RQ1\&2)}\label{h.acpoz1aemz1}

We show compliance results across experimental conditions in Table~\ref{tab:compliance}. Contrary to our initial expectations, the specific policy language (comprehensiveness and purpose) had little influence on compliance in experiment 1, where participants were asked to comply with the policy but compliance was not enforced through the grading rubric.  A few results are instructive though: Participants across all policy conditions were likely to share model accuracy (\textcircled{\fAccurate}), even though it was only required in the comprehensive policy. For other requirements only stated in the comprehensive policy, such as identifying model limitations (\textcircled{\fMisuse}), we saw slightly higher compliance when the requirement was actually stated, but only marginally so. In contrast, participants rarely identified the responsible organization to contact in case of harm (\textcircled{\fResponsible}) even though this was stated in the first sentence of the policy. This suggests that participants largely wrote what they understand and what is intuitive to them, often ignoring (or failing to comply with) other parts of the policy.  Regarding policy purpose, we did not detect any differences, quantitatively nor qualitatively. This lack of differences across policy comprehensiveness and purpose is why we did not vary policy language in the second experiment.

In experiment 2, we tightened policy enforcement with a stricter grading rubric. We found that observed compliance increased for almost every policy requirement (statistically significant except \textcircled{\fResponsible},\footnote{ The ``organization responsible'' requirement \textcircled{\fResponsible} was not included in the second experiment. The lack of change here supports the finding that the other changes are due to the treatment effect (enforcement).} \textcircled{\fClear}, and \textcircled{\fEight}; see appendix \gencite{1jmMl}{(Appendix: Policy Alone Is Probably Not the Solution: A Large-Scale Experiment on How Developers Struggle to Design Meaningful End-User Explanations 2025)}{}), such as including limitations of use (\textcircled{\fMisuse}), and reporting of misuse(\textcircled{\fReporting}).  Still, compliance was still fairly low for several requirements that did not align easily with technical explainability tooling, such as reporting personal data used (\textcircled{\fPersonal}, 20\%), and reporting path for misuse and harm (\textcircled{\fReporting}, 47.1\%). Low compliance, even despite enforcement, suggests that participants did not understand the requirements or were not able to comply. For example, participants sometimes encouraged users to report issues, but without providing a concrete reporting process or contact point. This explains also the lack of difference in complying with plain language requirements\textcircled{\fClear} and \textcircled{\fEight}), as almost all participants failed at this; only 4 individual explanations passed our assessment of 8th grade reading level, and not a single solution passed this for both global and individual explanation. 
\figureA

Analyzing participants' (often cursory) self-assessments of compliance in the first experiment, we found that participants often claimed that they complied even though they quite obviously did not. A notable only exception was that many participants acknowledged that they did not know how to comply with writing accessible explanations (\textcircled{\fEight}). In reflections, almost every participant described difficulty writing clear and accessible explanations \gencite{pjxS}{(Nahar et al. 2024)}{}. Some participants described this task as potentially insurmountable, like SP12: ``\emph{The requirement to use plain language can be at odds with the complexity inherent in automated systems, particularly in AI and machine learning models.}'' The necessity and trickiness of balancing was a common theme, and some participants thought they had done acceptably given resource constraints. For example, SP113 argued, \emph{``Fully complying with the policy can also take up a lot of extra time and cause stress. Engineers should be spending more time working on actual systems than writing up documentation […] perfect English and documentation skills aren't typically required of software experts}.'' Ultimately, some participants recognized that they were falling short in the requirement to write clearly but were unable to come up with a good solution. 

\subsection{Explanations were not meaningful for their intended end users (RQ2).}\label{h.wagc3vai72ns}

No explanation was fully compliant with policy; in particular, participants failed the requirement of a clear and accessible explanation. Even when participants complied, we often noticed shallow solutions that technically complied with the language of the policy, but did little to further the policy's purpose in our judgment, suggesting a check-the-box approach to compliance rather than careful engagement. We judged almost all explanations as likely incomprehensible to the intended users, and thus as ineffective. To better understand the problems beyond compliance, we thus analyzed common problems in an open-ended fashion (cf. Sec. 3), resulting in four failure modes we discuss here. In Table~\ref{tab:failure-modes-detailed}, we report the failure modes for global and individual explanations across both experiments (as policy conditions in the first experiment made no meaningful difference, we group them together here; details in the appendix \gencite{1jmMl}{(Appendix: Policy Alone Is Probably Not the Solution: A Large-Scale Experiment on How Developers Struggle to Design Meaningful End-User Explanations 2025)}{}). 
\figureFailure

\subsubsection{\faRobot \ Failure mode 1: Inscrutable or technically incorrect explanations}\label{h.m4mz0uip5kph}
Some explanations failed at basic intelligibility such that even experts cannot reasonably interpret what is shown. Several of these cases reflected misunderstandings of how explainability tools should be applied or interpreted. Common errors included presenting raw or opaque artifacts (e.g., arrays of pixel values, unlabeled SHAP outputs, or internal CNN activations), omitting essential context (e.g., feature names, scales, or reference points), or producing technically incorrect explanations due to misuse of explainability libraries (see appendix \gencite{1jmMl}{(Appendix: Policy Alone Is Probably Not the Solution: A Large-Scale Experiment on How Developers Struggle to Design Meaningful End-User Explanations 2025)}{}). For example, SP2 printed the numerical SHAP values of the pixels of only the top row of the image as an array of numbers (see appendix \gencite{1jmMl}{(Appendix: Policy Alone Is Probably Not the Solution: A Large-Scale Experiment on How Developers Struggle to Design Meaningful End-User Explanations 2025)}{}), offering no visual or semantic grounding and FP7 presented a SHAP waterfall plot attributing a prediction to individual image embedding dimensions (e.g., \emph{image\_embedding\_973}). This failure mode was relatively uncommon (7\% of individual explanations and 12\% of global explanations), and stricter enforcement in the second iteration did not meaningfully change these rates.

\subsubsection{\faUserCog \ Failure mode 2: Understanding explanations requires ML expertise}\label{h.fy1uylp1dnr3}
We judged 88\% of all solutions as likely inscrutable to end users without ML expertise. Participants frequently presented explanations in a disjointed, fragmented, check-the-box manner, lacking a coherent form aligned with the designated stakeholders. Very commonly, explanations resembled internal documentation intended for technically competent peers (e.g., machine learning experts). For instance, in almost half of the individual explanations, participants provided images that highlight areas without describing the significance of those areas (see Fig.~\ref{fig:overview}). Furthermore, explanations commonly included jargon-heavy language,  such as \emph{kappa}, \emph{confusion matrix}, or \emph{train/test data} instead of domain-appropriate medical language such as sensitivity, specificity, or efficacy or plain language descriptions for lay users. These explanations make sense to the developer, but are difficult to follow for anyone not immersed in the same exercise or knowledge base. Even when participants sometimes attempted to translate technical concepts into plain language, they provided lengthy descriptions of \emph{surrogate models} (FP14) or the concept of \emph{feature importance} (FP7), that are likely not relevant to patients' information needs. Even though plain and accessible language was required by the policy, even stronger enforcement did not improve these problems.

\subsubsection{\faUserMd \ Failure mode 3: Understanding explanations requires medical expertise }\label{h.uncnbclvf09r}
In 22\% of the individual explanations intended for patients, the explanation relies on medical terminology (e.g., \emph{neovascularization}, \emph{microaneurysms}, or the \emph{peripapillary region}) without additional context that is unlikely to be intelligible to patients without medical training. This was again largely unaffected by enforcement (not statistically significant; see appendix \gencite{1jmMl}{(Appendix: Policy Alone Is Probably Not the Solution: A Large-Scale Experiment on How Developers Struggle to Design Meaningful End-User Explanations 2025)}{}). Since global explanations were intended for nurses, we accepted such language there.

\subsubsection{\faProjectDiagram \ Failure mode 4: Explanations failed to consider the larger context and purpose of the AI system}\label{h.3yv5aa8kwj2h}
Many participants (66\%) failed to embed explanations in the context of a larger system or use where it is used as part of a workflow. For example, some global explanations reported model accuracy by subpopulation, but did not highlight those subpopulations as ones that should be approached with care in the text for healthcare professionals. Only a few explanations for patients included information about ``what does this mean for me'' or ``what are next steps.'' These solutions did not consider explainability as one contribution to a larger sociotechnical system aimed at reducing patients' risk of blindness. Embedding explanations in system context and purpose makes the tool more useful, and is especially critical in healthcare settings \gencite{WBqDh,3x0TR}{(Wang et al. 2023; Lebovitz et al. 2022)}{}. Notably, tightening policy enforcement was associated with an improvement regarding this failure mode, especially in global explanations (the only statistically significant result in failure modes analysis). Here, even check-the-box compliance required some engagement with harms, mitigations, and reporting, that go beyond a narrow focus on the model.

\subsubsection{Explanations suitable for end users}\label{h.kr7pqehiii6u}
While the vast majority of explanations were not plausibly targeted to patients or nurses, some participants did offer explanations that we thought were plausibly targeted to those end users. ``Good'' global explanations contained information presented in a clinically useful way (e.g., in terms of false positives and false negatives), showcasing the limitations and biases of the model to spur humans to challenge the model's results when it would matter the most for patient outcomes. We did not necessarily expect training data information or technical model details to be included, which is required for regulatory U.S. Food and Drug Administration clearance of medical devices, but usually not included in practitioner handbooks.

Properly targeted individual explanations used clear and accessible language, employing visuals and describing what they showed. They clearly marked the predictive result and posed and answered the question of ``what does this mean for me?'' For instance, after listing the patient name/ID, gender, age, and diagnosis on separate lines, the FP01 offered the following summary text: \emph{``Your eye scan shows proliferative diabetic retinopathy, a serious condition. This involves the growth of new, abnormal blood vessels in your retina, which can lead to severe vision impairment or blindness. Please seek urgent medical attention from an eye care specialist.''} Generally though all solutions that were tailored to patients and avoided the failure modes above still included way more information than clinical professionals that we spoke to preferred -- for example, explaining how to read the annotated image from an explainability tool, rather than omitting such visualization or merely providing reference images of diabetic retinopathy at different stages for the individual to compare to their own image. Norms of clinical communication \gencite{p8Hx}{(Menon et al. 2024)}{}, that include only the prediction, the personal data used (to comply with the policy requirement), what the patient should do next, and directing the patient to a number or organization if they had questions or were concerned about the accuracy of the result. 

\section{Discussion and Conclusion}\label{h.kk4uyqlri15k}
Given the minimal training and guidance provided, we did not expect participants to deliver high quality explanations appropriate for patients or nurses. We had some reason to expect that the policy would make some difference on explanations, by providing symbolic guidance as an incentive to nudge developers new to these ideas toward better explanations with a clear purpose and audience or whether the threat of legalistic sanction would propel them toward at least check-the-box compliance. We hoped that participants would realize how challenging it was to provide explanations that were suitable for end users. And given the same time frame and the same basic education on explainability as everyone else, some were able to provide end user-targeted explanations, suggesting it is not an impossible expectation. But we found little evidence that policy and increased enforcement substantively affected explanation quality. 

Ultimately, most participants did not seem to grasp how misaligned their explanations were for the needs of end users; they failed to understand recipients' information needs as distinct from their own. In their reflections, they chiefly reported their struggle to convey information in eighth grade language, finding it frustrating and onerous. They rarely discussed difficulty in interpreting the policy or ambiguity of terms (e.g., what ``dignity'' might mean, or differences in the expertise of different humans that the AI might collaborate with). They also did not acknowledge the difficulty of knowing what a patient or nurse would want to have explained. Though we anticipated that the policy purpose might guide participants in what information to provide in the first experiment, it had no recognizable influence. The symbolic, normative level of policy had little impact.

When we emphasized the legalistic side of the policy by increasing enforcement and threat of sanction in the second experiment, participants made fairly incremental changes to explanations. The effect of just providing a policy without the more heavyweight infrastructure of traditional software certification regimes (e.g., trainings, consultants, implementation guidelines) was limited, as visible in the observation that explanation quality was still low, even when compliance went up when enforced in the second experiment. With increased policy enforcement, we observed that participants engaged with some concerns beyond the model, for instance, including next steps for patients in individual explanations or guidelines for nurses on the quality of images for the tool. However, it did little to foster deeper engagement and perspective taking, and solutions still contained too much technical jargon and too much information.

We see the main failure of most solutions rooted in a lack of understanding user needs and a lack of perspective taking, matching the trope of empathy-challenged engineers not equipped to design user experiences for others \gencite{hptf}{(Cooper 2004)}{}. While the policy mandates clear and accessible language, our participants did not know how to approach this without dedicated instruction or access to experts. Like most engineers in practice, our participants studied engineering topics and not how to become a great writer or UX designer. Generally, participants largely followed their own intuition and focused on aspects of the explanation that seemed already most familiar to themselves (e.g., explanation tools for image models). The policy by itself stated policy goals and outcome requirements, but did not suggest how to get there (e.g., perspective taking, personas, interviews \gencite{509Q}{(Hanington and Martin 2019)}{}). Our participants did not know how to get there on their own, and it may seem unfair to expect them to do it when asked without prior training. If we want to shift responsible engineering practices with lightweight interventions, this will be an important obstacle to overcome.

\subsection{Toward better end-user explanations:  Recommended interventions}\label{h.13t88wphbjw8}
Our experimental results and findings about low quality explanations establish a baseline for other interventions and future research. It seems clear that policy guidance alone, even combined with enforcement (in the form that we explored, without a heavyweight regulatory framework) is unlikely to move the needle much and other, possibly complementary, interventions are necessary. In what follows, we draw on our findings to offer suggestions for policymakers and educators. 

\subsubsection{Recommendations for policymakers}\label{h.w16633hezjuz}
Our experiments show that policy, with compliance enforced, has some effect on explanations, but is insufficient to ensure ``good'' explanations. Our experiments suggest the gap between policy ideals and developers' sense and ability remains large, and more work is necessary to close it. The experiments illustrate that there is a potentially large disjuncture between policy on the books (as in the language used in the EU AI Act and in in-house responsible AI policies) and the interpretations made by developers, and that it may be unfair and unwise to leave it to developers to fill this gap. 

While there will be intermediaries, including compliance experts, to help bridge the gap, policymakers should consider the different needs of various stakeholders in writing policy, aiming explicitly to provide guardrails for innovation \gencite{oT4B}{(Suran M 2024)}{}. To make this possible, policymakers should give more guidance to developers to translate the intent of the policy, possibly down to the level of concrete suggestions for techniques and processes (e.g., interviews and personas for design, controlled experiments to evaluate effectiveness). Alongside this, training developers on how to demonstrate compliance is necessary -- and what would be considered adequate evidence and not just simply check a box with minimal effort. It would also be worth exploring how to more deeply instill a mission of the policy purpose in developers, which seemed entirely ignored in our experiment. Much work remains to be done to identify effective mechanisms of guidance and evaluation (e.g., auditing, certification) to ensure actual engagement with policy goals. Altogether, we maintain that this will result in more concrete policies that will provide actionable guidance to software developers and regulators to evaluate system qualities.

\subsubsection{Recommendations for educators}\label{h.y89yi21j55ta}
Findings from the experiments concerning developer education about explainability have pedagogical relevance inside the classroom and beyond in corporate training, online materials, and self-learning. First, we encourage instructors to \emph{engage student developers in critique and revision to improve explanations.} Instructors (or LLMs) can model and guide students through writing strategies. Following established pedagogical methods for cognitive process theory, which guides many writing classes \gencite{keYcC,3vc7p,t3XGu,kFqQw}{(Flower and Hayes 1981; Flower 1981; Beard et al. 2020; Ericsson 2017)}{}, instructors should help students list initial goals for explanations, then point out the ones that are in tension with one another. After a first draft, students should be asked to revisit and revise them. Instructors should assign students different stakeholders, and then in class, compare and discuss the explanations by stakeholder type to underscore their different needs. Assignments should ask students which explainability techniques advance which goals, encouraging students to reflect on their choice and use of explainability techniques (\emph{``how does using SHAP address your specific sub-goal?''}) as well as the construction of the text making up explanations (\emph{``tell me how you were thinking about your end user when you decided on this word choice''}). This makes students' justifications more explicit and defined in their own minds. These techniques and strategies should be used in combination.

Second, instructors can \emph{emphasize the domain and end user in teaching explainability techniques.} Research has shown the effectiveness of real-world examples, like site visits of clinics, watching a video about the context of use, and interviewing stakeholders about their needs to instill a sociological imagination \gencite{oICs5,DBTlY}{(Dowell 2006; Olsen 2016)}{}. Instructors should discuss the historical, cultural, and social elements of the assignment scenario, and invite discussion of which explainability techniques fit best within the domain and why, outlining alternative interpretations \gencite{aHwIO,Zj2SC}{(Robert J. Hironimus-Wendt and Lora Ebert Wallace 2009; Hirshfield 2022)}{}.

In our setting, participants could have benefited from the interaction with clinical practitioners or affected patients, or at least from the creation of personas \gencite{hptf}{(Cooper 2004)}{} for nurses and patients. Ideally, developers should test their explanations on an end user (or at least a chatbot stand-in). This active learning on test patients is a concept well explored in medicine, and we can learn from how a culture of careful end-use explanations is crafted in the context of clinical communication (e.g., a doctor explaining a diagnosis to a patient) \gencite{p8Hx}{(Menon et al. 2024)}{}. Clinical communication is both regulated but also actively taught:  ``standardized patients'' following a script interact with medical students to help them practice and improve their clinical assessment and communication skills \gencite{CxtsW,QLKV2,Ki4nj}{(Spencer et al. 2000; Kneebone et al. 2006; Bokken et al. 2010)}{}. In this manner, doctors are taught to anticipate patient perspectives. These curriculum innovations establish norms and practices beyond regulatory requirements. We propose a similar dual focus on design and establishing norms in our vision for pedagogy. We expect that the insights obtained from an HCI design course would aid developers in building these perspectives.

\subsubsection{Opportunities for tooling}\label{h.ajz6uvhdguln}
To shift norms of responsible engineering with lightweight interventions beyond education, LLMs and chatbots provide new opportunities too. Rather than relying on the flawed self-assessment of practitioners (our participants were really bad at recognizing their own mistakes), with custom prompting and some calibration LLMs can provide some initial critiques of explanations following an assessment rubric -- as we have explored when coding our participant's solution. Our results about common failure modes could also be used to build analysis tools that detect these. In addition, LLMs are now increasingly used to create personas and to interact with them \gencite{L4Q6,AI29,uEqC}{(Prpa et al. 2024; Schuller et al. 2024; Choi et al. 2024)}{}; developers could use them for initial interviews to identify explanation needs \gencite{qJez}{(Lojo et al. 2025)}{}, to force a perspective shift, possibly recognizing a gap in understanding, that then triggers subsequent exploration (or outreach to experts). Finally, there is room to provide tooling to easily create prototypes of explanations, so that both developers and interviewed users gain a better sense of what is possible, to explore a wider design space, rather than following what is already intuitive to them. How to build such tools and embed them in a process such developers appreciate them and engage deeply with them, rather than checking boxes when these tools are forced on them is a continues design challenge, with many ideas from process integration, to champion-models, to gamification, to marketing strategies explored in the literature \gencite{hAo6,enga,PpMe,Ogd1,p7FS,5v7o,4Ysh}{(Nahar et al. 2025; Ballard et al. 2019; Deng et al. 2023; Crampton 2021; Howard and LeBlanc 2003; Bhat et al. 2023; Kim et al. 2025)}{}.

\end{document}